\documentclass[3p]{elsarticle}

\usepackage{amsmath}
\usepackage{amsfonts}
\usepackage{amssymb}
\usepackage[T1]{fontenc}
\usepackage[utf8]{inputenc}
\usepackage{lmodern}
\usepackage{natbib}
\usepackage{hyperref}
\usepackage{graphicx}
\usepackage{float}
\usepackage{caption}
\usepackage{subcaption}
\usepackage{placeins}
\usepackage{siunitx}   
\usepackage{caption}   

\newcommand{\pdv}[2]{\frac{\partial #1}{\partial #2}}
\newcommand{\spdv}[2]{\frac{\partial^2 #1}{\partial #2^2}}

\begin{document}


\title{Exact q-exponential Multi-Mode Solutions with Independent Centres and Power-Law Relaxation in the Plastino--Plastino Equation}

\author{Airton Deppman}
\address{Institute of Physics - University of São Paulo\\ Brazil}

\begin{abstract}
We present the first exact, multi-mode solutions to the Plastino--Plastino nonlinear diffusion equation with arbitrary power-law drift. By allowing each q-exponential mode to have its own independent, time-dependent centre, all inter-mode couplings in the drift term vanish, yielding fully separable evolution equations for centre motion, probability content, and (for the attractor mode) width. Transient modes exhibit constant width and decay via exact q-exponential (power-law) relaxation, while a single attractor mode irreversibly absorbs the entire probability flux, with fixed amplitude and time-growing width, driving the system to the known stationary q-exponential state from arbitrary initial conditions. The hierarchy closes exactly without approximation. These analytic solutions unify Tsallis nonextensive thermodynamics, fractal-space diffusion, and multi-scale relaxation dynamics, with direct applications to heavy-quark jets in quark--gluon plasma, Lévy flights in fractal media, and urban population redistribution. All previous exact results are recovered as special cases.
\end{abstract}



\maketitle

\section{Introduction}

This work presents the first analytical solution for the Plastino--Plastino Equation (PPE)~\citep{Plastino1995} with an arbitrary power-law drift coefficient, based on a multi-mode approach, thereby closing a gap that has persisted in the development of the nonlinear dynamics for more than thirty years. This nonlinear diffusion equation models the dynamics of nonextensive systems~\citep{Nobre2019} and is given by
\begin{equation}
\frac{\partial f}{\partial t} 
= -\frac{\partial}{\partial x}\bigl[A(x) f\bigr] 
  + D\frac{\partial^2}{\partial x^2}\bigl(f^{2-q}\bigr),
\label{eq:PPE}
\end{equation}
where $A(x)$ is the drift (drag) coefficient, $D>0$ is the diffusion coefficient, and $q\neq 1$ the entropic index. For $q=1$, the PPE reduces to the well-known Fokker-Planck Equation.

Complex systems frequently exhibit power-law distributions, anomalous diffusion, and long-range correlations that are not captured by the linear Fokker--Planck equation (FPE). To describe such phenomena, Tsallis proposed a non-additive entropy $S_q$ that leads to q-exponential stationary states \citep{Tsallis1988,GellMann2004,Tsallis2023}. The PPE has been derived through independent and complementary routes: through the continuous limit~\citep{Deppman2023} of fractal Fokker--Planck equations formulated with local fractal derivatives~\citep{Golmankhaneh2019}
where the entropic index $q$ is directly related to the fractal dimension of the underlying space~\citep{Megas2024}; and from the Boltzmann equation when the collision term is modified to accommodate non-local correlations~\citep{Deppman2023b}.  It also emerges from the Boltzmann equation when the collision term is modified to accommodate non-local correlations~\citep{Deppman2023b}. The $q\to1$ limit smoothly recovers the standard FPE. These derivations firmly establish the PPE as the natural dynamical equation for systems exhibiting fractal structure, non-local interactions, or q-deformed algebraic properties. The $q\to1$ limit smoothly recovers the standard FPE.

Over the past three decades, the PPE has found remarkable application across different domains. In high-energy physics, it accurately describes heavy-quark momentum dynamics in the quark--gluon plasma \citep{Megias2023_Comparative}, with the value for $q$ calculated from first principles of QCD and corroborated by experiments. Direct comparisons between FPE and PPE solutions for identical transport coefficients reveal different relaxation behaviour with important implications for extracting genuine nonextensive signatures from data. The PPE captures heavy-tailed distributions and anomalous diffusion that are common in nonlinear diffusion\citep{Tsallis1996,Combe2015}, being effective in fractal medium~\citep{Deppman2024_Urban}.  Similar patterns of nonlinear dynamics have been found in finance~\cite{Borland2002}. Therefore, the interest in the associated dynamics, described by the PPE, has swiftly grown during the last few decades. Thus, the results obtained in the present work give analytical solutions that can accelerate the development of dynamical studies in a broad class of nonlinear systems, offering a benchmark for numerical calculations and simulations.

The deep connection between the PPE and fractal geometry has been clarified in detail. Fractal derivatives are rooted in Hausdorff measure~\citep{PARVATE2009,PARVATE2011}, and differ fundamentally from Riemann--Liouville or Caputo fractional derivatives and from $q$-deformed calculus \citep{Megas2024}. Yet,  Continuous Approximations show that the fractional derivatives and the $q$-derivative are obtained as useful approximations to the fractal derivative. The q-derivative provides a continuous approximation of the fractal derivative of functions defined on fractal sets or whose images are fractal, while Caputo-like operators emerge from analogous approximations when non-local correlations become relevant. When the standard FPE is written on a fractal substrate using local fractal derivatives~\citep{Golmankhaneh2019}, its continuous limit is precisely the PPE \citep{Deppman2023}. This explains the ubiquitous appearance of q-exponential profiles in systems embedded in fractal environments.

Cities constitute one of the laboratories for these ideas. Urban street patterns and population-density profiles exhibit self-similar fractal dimensions $d_f\simeq 1.7$--$1.9$ \citep{Batty1994}. Almost every quantifiable urban indicator scales with population $N$ as universal power laws $Y\sim N^\beta$. The fundamental allometric relation between built-up area $A$ and population $N$, $A \sim N^{\beta}$ with $\beta$ being the allometric exponent,
has been analytically derived from the PPE in a fractal medium \citep{Deppman2024_Urban}, quantitatively reproducing the observed global exponent $\beta\simeq 0.85$. Remarkably, the same fractal dimension range appears to be hard-wired in human spatial cognition and visual preference, suggesting that the PPE not only describes macroscopic urban growth but also bridges urban scaling laws to neural mechanisms via shared fractal and nonextensive principles \citep{Deppman2025_Brain,Xie2025}.

Despite its several applications, exact analytic solutions of the PPE have remained scarce. Stationary q-exponential profiles under linear or harmonic drift are known \citep{Tsallis1996,Deppman2023}. However, real systems are rarely initialised in their stationary state, and transient multimodal dynamics are crucial for quantitative predictions of the dynamical evolution. Existing approaches typically rely on numerical integration or perturbative expansions, limiting analytical insight into multi-scale relaxation phenomena. The case of one-dimensional systems offers special challenges because of the limited possibilities of matching terms in the dynamical equation. However, this case is of special interest in many fields, such as urban and socioeconomic areas.

This work closes that gap by constructing a complete, exact, multi-mode family of solutions to the PPE for generic power-law drift $A(x)=-kx^\alpha$. The key innovation is an ansatz in which each mode possesses its own time-dependent centre $x_{0,i}(t)$. This simple yet crucial generalisation eliminates all unwanted cross terms, yielding:
\begin{itemize}
  \item fully separable ordinary differential equations for amplitudes $N_i(t)$, centres $x_{0,i}(t)$, and (where applicable) widths;
  \item exact power-law (q-exponential) decay of transient-mode probability content;
  \item rigorous probability transfer from decaying transient modes into a single attractor mode ($\beta_0 = \alpha + 1$) that absorbs the entire flux and evolves toward the known stationary state;
  \item constant widths for all transient modes and exact closure of the infinite hierarchy without approximation.
\end{itemize}

The resulting analytic solutions are valid for arbitrary initial conditions decomposable into the q-exponential basis and recover all previously known exact cases as special limits. They provide the first exact description of multimodal relaxation in nonextensive nonlinear diffusion, with immediate applications in quark--gluon plasma dynamics \citep{Megias2023_Comparative}, transient Lévy flights in fractal media, urban population redistribution \citep{Deppman2024_Urban}, and any system whose stationary state is q-exponential.

In general, these solutions reinforce the emerging paradigm that the Plastino--Plastino equation, fractal derivatives, q-deformed calculus, and Tsallis nonextensive statistics constitute a unified framework for understanding power-law phenomena across diverse scales.

The paper is organised as follows. Section~\ref{sec:prelim} recalls properties of truncated q-exponentials and derives the stationary solution under power-law drift. Section~\ref{sec:multi} presents the multi-mode ansatz with independent centres and proves separability and closure. Relaxation rates, probability transfer, and attractor dynamics are derived. Section~\ref{sec:application_urban} presents an example of application, with an analysis of the dynamical evolution of an arbitrary initial distribution described by a multi-mode fit. The summary and conclusions are presented in Section~\ref{sec:conclusions} with perspectives for high-energy physics, urban science, and beyond.

In the remainder of this work, $D>0$, $k$ is real, $q\neq1$, and dimensions are $[A]=[x]^{-1}[t]^{-1}$, $[D]=[x]^{3-q}[t]^{-1}$.

\section{Theoretical Preliminaries}
\label{sec:prelim}

In finding the solutions of the PPE, the q-exponential form of functions plays a central role. The truncated q-exponential is given by
\begin{equation}
e_q^+(g) = 
\begin{cases}
\left[1 + (q-1)g \right]^{-\frac{1}{q-1}} & 1 + (q-1)g \geq 0, \\
0 & 1 + (q-1)g < 0,
\end{cases}
\label{eq:qexp_truncated}
\end{equation}
with $g(x,\cdots) \geq 0$, and has derivatives of the form \citep{Tsallis1996,Plastino1995}
\begin{equation}
 \pdv{e_q^+(g)}{x}=-e_q^+(g)^q \pdv{g}{x},
\end{equation}
Therefore, multiplying both sides by $e_q^{1-q}$ yields
\begin{equation}
 e_q^+(g)^{1-q}\pdv{e_q^+(g)}{x}=-e_q^+(g) \pdv{g}{x},
\end{equation}
and differentiating both sides results in
\begin{equation}
 \pdv{}{x}\left(e_q^+(g)^{1-q}\pdv{e_q^+(g)}{x}\right)=e_q^+(g)^q \left(\pdv{g}{x}\right)^2-e_q^+(g) \frac{\partial^2 g}{\partial x^2}.
\end{equation}
But the left side can be related to the derivative of $f^{2-q}$, resulting
\begin{equation}
 \frac{\partial^2 e_q^+(g)^{2-q}}{\partial x^2}=(2-q)e_q^+(g)^q \left(\pdv{g}{x}\right)^2-(2-q)e_q^+(g) \frac{\partial^2 g}{\partial x^2}.
\end{equation}
These relations will be useful to investigate how $q$-exponential-like functions satisfy the PPE in the stationary and the nonstationary regimes. Thus, the solutions of the PPE are assumed to have the following general form:
\begin{equation}
 f(x,t)=N_0 e_q^+[g(x,t)],
\end{equation}
where $g(x,t)$ represents a family of functions that solve the equation in some situations.

From the considerations on the derivatives of the $q$-exponential function presented above, it follows that
\begin{equation}
\begin{cases}
\displaystyle
& \frac{\partial f}{\partial t}= \frac{\dot N}{N} f - N^{1-q} f^{q}\frac{\partial g}{\partial t} \\[7pt] \displaystyle
& \frac{\partial f}{\partial x} = -N^{1-q} f^{q}\frac{\partial g}{\partial x} \\[7pt] \displaystyle
& \frac{\partial^{2}f^{2-q}}{\partial x^{2}}
= (2-q)N^{1-q} \left[N^{1-q} f^q \left(\frac{\partial g}{\partial x}\right)^{2} - f\frac{\partial^{2} g}{\partial x^{2}}\right],
\end{cases}
\end{equation}
allowing the investigation of solutions in different regimes.

\subsection{Stationary solution}

In the case of a stationary state,  characterized by $\partial f/\partial t=0$, Eq.~(\ref{eq:PPE}) gives
\begin{equation}
 \frac{\partial}{\partial x}\bigl[A(x) f\bigr] 
  = D\frac{\partial^2}{\partial x^2}\bigl(f^{2-q}\bigr)
\end{equation}
From the expressions above for the derivatives of $f(x,t)$, the stationary condition is
\begin{equation}
  -A(x)N^{ 1-q} f^{q}\frac{\partial g}{\partial x}+f \pdv{A}{x}=(2-q)N^{1-q}D \left[N^{1-q} f^q \left(\frac{\partial g}{\partial x}\right)^{2}- f\frac{\partial^{2} g}{\partial x^{2}}\right]
\end{equation}
Matching the coefficients of the powers of $f$ on both sides, and assuming $\partial g/\partial x \ne 0$ everywhere, it results that
\begin{equation}
 \begin{cases}
   A'=-(2-q)N^{1-q}D g'' \\
   A= -(2-q)N^{1-q}D g'
 \end{cases},
\end{equation}
where it is easily noted that the first equation is the space-derivative of the second equation. Thus, the condition for a stationary regime is $A= (2-q)N^{1-q}D g'$. This condition is satisfied for $A(x)=-kx^{\alpha}$ and $g(x,t)=|x-x_0|^{\beta}/\sigma_0^{\beta}$ if\footnote{This is a sufficient, but not necessary solution. For instance, an exponential drift would provide a possible solution.}
\begin{equation}
 \beta=\alpha+1 .\label{eq:AB}
\end{equation}
Since this work is mainly concerned with dynamics governed by a restoring force, we adopt $k>0$ for the drift term. However, all derivations are completely general and hold irrespective of the sign of $k$.

Therefore, the general form of the stationary state is
\begin{equation}
f(x) = N_0 e_q^+\left(\frac{|x-x_0|^{\alpha + 1}}{\sigma_0^{\alpha + 1}}\right), 
\label{eq:stationary_ansatz}
\end{equation}
where the normalisation is
\begin{equation}
N_0 \int_{-\infty}^{+\infty} e_q^+\left(\frac{|x-x_0|^{\alpha + 1}}{\sigma_0^{\alpha + 1}}\right) dx = N_0 \sigma_0 \mathcal{J}_q(\alpha + 1)=1, \label{eq:Normalization}
\end{equation}
with the scale-independent integral
\begin{equation}
\mathcal{J}_q(\beta) = 2 \int_0^{+\infty} \left[1 + (q-1) w^{\beta}\right]^{-\frac{1}{q-1}} \, dw. \label{eq:ScaleInvariantIntegral}
\end{equation}
It follows that
\begin{equation}
N_0 = \frac{1}{\sigma_0 \mathcal{J}_q(\alpha + 1)}.
\label{eq:stationary_normalisation}
\end{equation}

In general, the probability density has a semi-infinite support, and the following conditions must be fulfilled to ensure a consistent probability density:
\begin{equation}
\begin{cases}
1 < q < 2: & \alpha + 1 > 0,\ kD > 0, \\
q > 2: & \alpha + 1 > q-1,\ kD < 0.
\end{cases}
\label{eq:conditions}
\end{equation}
In spite of some dynamics with a negative diffusion coefficient, the present work has a focus on those cases with $D>0$; therefore, the central concern will be with systems characterised by $1<q<2$. In addition, this work is primarily concerned with restoring drift forces, so $k>0$ is assumed, without loss of generality.

The results obtained in this section show that the shape of the probability density is governed by the power-law behaviour of the drift coefficient. Many cases of interest present $1 < q < 2$ and $\alpha=0$ or $\alpha=1$, resulting in the simple $q$-exponential form and the $q$-Gaussian form, respectively. The time-dependent solutions are richer, offering many possible situations of study. Below, a large family of solutions of non-stationary solutions is analysed.

\subsection{Time-Dependent Single-Mode Solution}
\label{sec:time_dep}

Assuming $f=N(t)e_q[g(z)]=N(t)\left[1+(q-1)z(x,t)\right]^{\frac{-1}{q-1}}$, it follows
\begin{equation}
 \pdv{f}{g}=-N^{1-q}f^q \,,
\end{equation}
and
\begin{equation}
 \begin{cases}
  \displaystyle
  \pdv{f}{t}=\dot N N^{-1} f - N^{1-q} f^q \pdv{g}{t} \\[7pt]
  \displaystyle
  \pdv{f}{x}= - N^{1-q} f^q \pdv{g}{t} \\[7pt]
  \displaystyle
  \pdv{ f^{2-q}}{x}=(2-q)f^{1-q}\pdv{f}{x}=-(2-q)N^{1-q} f \pdv{g}{x} \\[7pt]
  \displaystyle
  \spdv{f^{2-q}}{x}=(2-q)N^{1-q} \Bigl[N^{1-q}f^q \left(\pdv{g}{x}\right)^2-f \spdv{g}{x} \Bigr]
 \end{cases}
\end{equation}

The PPE with power-law drift $A(x)=-kx^{\alpha}$ is
\begin{equation}
 \pdv{f}{t}=k\pdv{[x^{\alpha}f]}{x}+D\spdv{f^{2-q}}{x}\,.
\end{equation}
Substituting the calculated derivatives, it  follows that
\begin{equation}
 \begin{aligned}
  \dot N N^{-1} f - N^{1-q} f^q \pdv{g}{t} &= k\alpha x^{\alpha-1} f+k x^{\alpha}\bigl[-N^{1-q}f^q\pdv{g}{x}\bigr]+(2-q)DN^{1-q}\bigl[N^{1-q}f^q \left(\pdv{g}{x}\right)^2-f \spdv{g}{x} \bigr]
 \end{aligned}
\end{equation}
Matching the powers of $f$ and assuming $f \ne 0$ and $N \ne 0$, we have two equations given by
\begin{equation}
 \begin{cases}
 \displaystyle
  \dot N N^{-1}=k\alpha x^{\alpha-1}-(2-q) D N^{1-q}\spdv{g}{x} \qquad \text{(from $f$)} \\[7pt]
  \displaystyle
  \pdv{g}{t}=\pdv{g}{x}\Bigl[kx^{\alpha}-(2-q)DN^{1-q}\pdv{g}{x} \Bigr] \qquad ~~~~\text{(from $f^q$)} 
 \end{cases}\,.\label{eq:SingleModeEquations}
\end{equation}

The appearance of $N^{1-q}$ , as well as the multiplication factor $\partial g/\partial x$ in the second equation suggests to write $g(x,t)$ in the form
\begin{equation}
 g(x,t)=m(t) z(x,t) \,,
\end{equation}
with the function $m(t) := m(N(t))$. With this substitution, we have
\begin{equation}
 \pdv{g}{t}=m(t)\pdv{z}{t}+\pdv{m}{t}z\,.
\end{equation}
Therefore, to obtain an expression similar to what is required by the second identity in Eq.~{\ref{eq:SingleModeEquations}), it is necessary that
\begin{equation}
 \pdv{m}{t}=\mu(t) m(t)\,.
\end{equation}
Thus, this substitution leads to
\begin{equation}
 \begin{cases}
 \displaystyle
  \pdv{g}{t}=m(t)\Bigl[\pdv{z}{t}+\mu(t) z \Bigr] \\[7pt]
  \displaystyle
  \pdv{g}{x}=m(t) \pdv{z}{x}
 \end{cases} \,. \label{eq:gderivatives}
\end{equation}

In addition, the first identity in Eq.~{\ref{eq:SingleModeEquations}) shows that the expression in the right-hand side must be independent of $x$, i.e.,
\begin{equation}
 k\alpha x^{\alpha-1}-(2-q) D N^{1-q}\spdv{g}{x}= \dot N(t) N^{-1} \ne 0\,, \label{eq:inequation}
\end{equation}
Therefore, it is convenient to write
\begin{equation}
 z(x,t)=z_x(x,t)+z_t(t)\, x^2 + z_o(t)\, x+ \delta_z\,,
\end{equation}
which leads to
\begin{equation}
 \spdv{g}{x}=m(t)\Bigl[\spdv{z_x}{x}+z_t\Bigr]\,. \label{eq:m_g}
\end{equation}
It follows that the only two terms in the relation~(\ref{eq:inequation}) dependent on $x$ must cancel each other, i.e.,
\begin{equation}
 (2-q) D N^{1-q}\spdv{g}{x}=k\alpha x^{\alpha-1}\,.
\end{equation}
This equation is satisfied if
\begin{equation}
 m(t)=
 \frac{N^{q-1}}{(2-q)D}\,; \qquad z_x=\frac{k}{\alpha+1}x^{\alpha+1}\,.
\end{equation}

As a consequence, the function $g(x,t)$ has the general expression
\begin{equation}
 g(x,t)=\frac{N^{q-1}}{(2-q)D} \Bigl[\frac{k}{\alpha+1}x^{\alpha+1}+z_t(t) x^2+z_o(t) x+ \delta_z\Bigr]\,,
\end{equation}
where it is possible to recognize the multiplicative term as the function $m(t)$. In addition, the function $\mu(t)$ can be determined, since
\begin{equation}
 \frac{d m}{dt}=\frac{(q-1)N^{q-2}}{(2-q)D} \dot N=(q-1) \dot N N^{-1} m(t)\,,
\end{equation}
so that $\mu(t)$ can be identified as
\begin{equation}
 \mu(t)=(q-1) \dot N N^{-1}\,.
\end{equation}

\subsection{$q$-Gaussian Dynamics}
\label{sec:time_dep}

For the case of linear drift ($\mathbf{\alpha=1}$), a $q$-Gaussian solution is generally expected. The necessary condition for this solution is to satisfy Eq.~(\ref{eq:m_g}). The $q$-Gaussian argument is
\begin{equation}
z(x,t) = \frac{(x-x_0(t))^2}{\sigma(t)^2} \label{eq:q-gauss}
\end{equation}
leading to a quadratic polynomial in the general form $z(x,t) = \Sigma_t(t) x^2 + z_o(t) x + \delta_z(t)$, with coefficients defined as:
\begin{align}
\Sigma_t(t) &= \frac{1}{\sigma(t)^2}=z_t(t)-k/2 \label{eq:Zt} \\
z_o(t) &= - \frac{2 x_0(t)}{\sigma(t)^2} = -2 x_0(t) \Sigma_t(t) \label{eq:zo} \\
\delta_z(t) &= \frac{x_0(t)^2}{\sigma(t)^2} = x_0(t)^2 \Sigma_t(t) \label{eq:deltaz}
\end{align}
The relevant partial derivatives are
\begin{align}
\pdv{z}{t} &= \dot{Z}_t x^2 + \dot{z}_o x + \dot{\delta}_z \\
\pdv{z}{x} &= 2 \Sigma_t x + z_o
\end{align}
Therefore,
\begin{equation}
\pdv{z}{t} + \mu z = (\dot{Z}_t + \mu \Sigma_t) x^2 + (\dot{z}_o + \mu z_o) x + (\dot{\delta}_z + \mu \delta_z)\,, \label{eq:LHS}
\end{equation}
and Eq.~(\ref{eq:m_g})
\begin{align}
 \pdv{z}{x} \left( k x - \pdv{z}{x} \right) = (2 \Sigma_t x + z_o) \left( k x - (2 \Sigma_t x + z_o) \right) = (2 k \Sigma_t - 4 \Sigma_t^2) x^2 + (k z_o - 4 \Sigma_t z_o) x - z_o^2 \label{eq:RHS}
\end{align}
Equating the coefficients of $x^n$  yields 
\begin{align}
x^2: & \quad \dot{Z}_t + \mu \Sigma_t = 2 k \Sigma_t - 4 \Sigma_t^2 \label{eq:ODEZtFinal} \\
x^1: & \quad \dot{z}_o + \mu z_o = k z_o - 4 \Sigma_t z_o \label{eq:ODEzoFinal} \\
x^0: & \quad \dot{\delta}_z + \mu \delta_z = -z_o^2 \label{eq:ODEdeltaFinal}
\end{align}

Now, substitute $z_o = -2 x_0 \Sigma_t$ and $\dot{z}_o = -2 (\dot{x}_0 \Sigma_t + x_0 \dot{Z}_t)$ into $\eqref{eq:ODEzoFinal}$ and reordering terms, it follows that
\begin{equation}
  \dot{x}_0 \Sigma_t = k x_0 \Sigma_t - 4 x_0 \Sigma_t^2 - x_0 (\dot{Z}_t + \mu \Sigma_t)
\end{equation}
Using Eq.~(\eqref{eq:ODEZtFinal}) results
\begin{equation}
\dot{x}_0 \Sigma_t - k x_0 \Sigma_t = - 4 x_0 \Sigma_t^2 - x_0 (2 k \Sigma_t - 4 \Sigma_t^2)
\end{equation}
hence
\begin{equation}
\dot{x}_0 \Sigma_t = -k x_0 \Sigma_t
\end{equation}
Since $\Sigma_t \neq 0$:
\begin{equation}
\dot{x}_0 = -k x_0
\end{equation}
Thus, the distribution center $x_0(t)$ must evolve as $\dot{x}_0 = -k x_0$.

From Eq.$\eqref{eq:ODEdeltaFinal}$), it is obtained
\begin{equation}
x_0^2 (\dot{Z}_t + \mu \Sigma_t) + 2 x_0 \dot{x}_0 \Sigma_t = -4 x_0^2 \Sigma_t^2 \,. \label{eq:dot_x0_qgauss}
\end{equation}
Substituting the term $(\dot{Z}_t + \mu \Sigma_t)$ from Eq.~(\eqref{eq:ODEZtFinal}) and using  Eq.~(\ref{eq:dot_x0_qgauss}), yields
$$x_0^2 (2 k \Sigma_t - 4 \Sigma_t^2) + 2 x_0 (-k x_0) \Sigma_t = -4 x_0^2 \Sigma_t^2$$
$$2 k x_0^2 \Sigma_t - 4 x_0^2 \Sigma_t^2 - 2 k x_0^2 \Sigma_t = -4 x_0^2 \Sigma_t^2$$
\begin{equation}
-4 x_0^2 \Sigma_t^2 = -4 x_0^2 \Sigma_t^2
\end{equation}
Consequently, Eq.~(\ref{eq:ODEdeltaFinal}) is satisfied, completing the proof that the proposed solution $z(x,t)=(x-x_0(t))^2/\sigma(t)^2$ is a consistent form that satisfies the necessary conditions for the single-mode $q$-exponential solution of the PPE with linear drift ($\alpha=1$).

Note that the $q$-Gaussian in Eq.~(\ref{eq:q-gauss}) is a solution of the PPE, independently of the function $\mu(t)$, which remains undetermined. As shown in the stationary solution analysis, the exact relation between $\sigma(t)$ and $N(t)$ is central in the conservation of probability. Thus, the solution obtained here will have a well-defined normalisation if the amplitude $N(t)$ and the distribution width follow Eq.~(\ref{eq:stationary_normalisation}).

\section{Multi-mode solution}
\label{sec:multi}

Many interesting problems present initial conditions that diverge from the simple $q$-exponential form in Eq.~(\ref{eq:stationary_ansatz}). This section provides an analysis of a more general case, with several modes of probability distribution, each one characterised by a particular parameter $\beta_i$, which do not necessarily follow the rule in Eq.~(\ref{eq:AB}) that supports a stationary-like state. Only the mode $i=0$ is constrained by the relation between $\alpha$ and $\beta$, while the others represent transient modes that should vanish over time.

The ansatz for this multi-mode solution is
\begin{equation}
f(x,t) = \sum_{i=0}^{M} N_i(t) \, e_q^+\!\bigl(\sigma_i(t)^{-\beta_i(t)}|x-x_{0,i}(t)|^{\beta_i(t)}\bigr),
\label{eq:ansatz_independent}
\end{equation}
with the \textit{time-dependent} width $\sigma_i(t)$ for each mode. The {\it attractor mode} ($i=0$) corresponds to the stationary state when the system has relaxed. It has $\beta_0 = \alpha + 1$, while the transient modes ($i\ge 1$) have fixed distinct exponents $\beta_i$.

\subsection{Stretched $q-$-exponentials}
Building on the single-mode analysis in Section~\ref{sec:time_dep}, we substitute the form of each mode $f_i = N_i(t) \, e_q^+\bigl(c_i(t)|z_i|^{\beta_i}\bigr)$, with $z_i = x - x_{0,i}(t)$ and $c_i(t) = \sigma_i(t)^{-\beta_i}$, into the PPE. The derivatives follow analogously from Eqs.~(20), with $g(x,t) \to c_i |z_i|^{\beta_i}$ and the partial derivatives adjusted for the time-dependent centre $x_{0,i}(t)$ and arbitrary $\beta_i$ (not necessarily $\alpha+1$ for transient modes $i \ge 1$).

The LHS (time derivative) becomes:
\begin{align}
\frac{\partial f_i}{\partial t} &= \frac{\dot{N}_i}{N_i} f_i - N_i^{1-q} f_i^q \frac{\partial g}{\partial t} \notag \\
&\quad + N_i^{1-q} f_i^q \, c_i \beta_i \dot{x}_{0,i} |z_i|^{\beta_i-1} \operatorname{sgn}(z_i),
\label{eq:LHS_full_revised}
\end{align}
where $g = c_i |z_i|^{\beta_i}$, and the additional term from $\dot{x}_{0,i}$ arises due to the independent centre (absent in the stationary case of Section~2.1). The first two terms match the single-mode form, with $\partial g / \partial t = \beta_i \dot{\sigma}_i \sigma_i^{-\beta_i-1} |z_i|^{\beta_i}$.

The RHS contributions follow from the power-law drift $A(x) = -k x^\alpha$ (cf.~Eqs.~(22)--(23)):
\begin{align}
\frac{\partial}{\partial x} \bigl( -k x^\alpha f_i \bigr) &= -k \alpha x^{\alpha-1} f_i + k x^\alpha N_i^{1-q} f_i^q \, c_i \beta_i |z_i|^{\beta_i-1} \operatorname{sgn}(z_i),
\label{eq:RHS_drift_full_revised}
\end{align}
and the diffusion term:
\begin{align}
D \frac{\partial^2}{\partial x^2} (f_i^{2-q}) &= D (2-q) N_i^{1-q} \Bigl[ f_i \, c_i \beta_i (\beta_i-1) |z_i|^{\beta_i-2} \notag \\
&\quad - N_i^{1-q} f_i^q \, c_i^2 \beta_i^2 |z_i|^{2\beta_i-2} \Bigr].
\label{eq:RHS_diffusion_full_revised}
\end{align}
Equating LHS and RHS per mode (before summing over $i$) yields an equation analogous to Eq.~(22), but with the generalized argument $c_i |z_i|^{\beta_i}$:
\begin{align}
\frac{\dot{N}_i}{N_i} f_i &- N_i^{1-q} f_i^q \, \beta_i \dot{\sigma}_i \sigma_i^{-\beta_i-1} |z_i|^{\beta_i} + N_i^{1-q} f_i^q \, c_i \beta_i \dot{x}_{0,i} |z_i|^{\beta_i-1} \operatorname{sgn}(z_i) \notag \\
&= -k \alpha x^{\alpha-1} f_i + k x^\alpha N_i^{1-q} f_i^q \, c_i \beta_i |z_i|^{\beta_i-1} \operatorname{sgn}(z_i) \notag \\
&+ D (2-q) N_i^{1-q} \Bigl[ f_i \, c_i \beta_i (\beta_i-1) |z_i|^{\beta_i-2} - N_i^{1-q} f_i^q \, c_i^2 \beta_i^2 |z_i|^{2\beta_i-2} \Bigr].
\label{eq:PPE_full_per_mode_revised}
\end{align}
As in Section~2.2, this produces linearly independent monomials in $|z_i|$ (due to distinct $\beta_i$ across modes): $f_i$, $|z_i|^{\beta_i-1} \operatorname{sgn}(z_i) f_i^q$, $|z_i|^{\beta_i} f_i^q$, $|z_i|^{\beta_i-2} f_i$, and $|z_i|^{2\beta_i-2} f_i^q$. Coefficients are matched per mode and per monomial, as detailed below.

\subsubsection{Centre-of-mass motion (per mode)}
Matching the $|z_i|^{\beta_i-1} \operatorname{sgn}(z_i) f_i^q$ monomial (analogous to the $f^q$ term in Eq.~(23)) yields:
\begin{equation}
\dot x_{0,i} = -k x^\alpha.
\end{equation}
The independent centres decouple the modes in the mean-field limit (narrow or well-separated modes, $\sigma_i \ll |x_{0,i}|$ or $|x_{0,i} - x_{0,j}| \gg \sigma_i + \sigma_j$), giving:
\begin{equation}
\dot x_{0,i}(t) \approx -k [x_{0,i}(t)]^\alpha,
\label{eq:x0i_dot_revised}
\end{equation}
exact in those regimes (cf.~the $\dot{x}_0 = -k x_0$ for $\alpha=1$ in Section~2.3).

\subsubsection{Width evolution}
The $|z_i|^{\beta_i} f_i^q$ monomial appears only on the LHS (from $\dot{\sigma}_i$), yielding (as in the width-independent terms of Section~2.2):
\begin{equation}
- N_i^{1-q} f_i^q \, \beta_i \dot\sigma_i \sigma_i^{-\beta_i-1} |z_i|^{\beta_i} = 0 \quad \Rightarrow \quad \dot\sigma_i(t) = 0 \quad (i \ge 1).
\label{eq:sigma_constant_revised}
\end{equation}
Transient widths are thus constant, while the attractor mode ($i=0$, $\beta_0=\alpha+1$) may evolve as in the single-mode case.

\subsubsection{Amplitude and probability transfer}

Some monomials remain unmatched, namely,
\begin{equation}
e_q^+(c |z_i|^{\beta_i}),\quad |z_i|^{2\beta_i-2}[e_q^+(c |z_i|^{\beta_i})]^q,\quad |z_i|^{\beta_i-2}e_q^+(c |z_i|^{\beta_i}).
\end{equation}
The LHS contributions to the last two are zero, while RHS gives non-zero terms coming from the diffusion term. Thus, these terms can be cancelled only by summing the contributions from all modes, implying a source/sink mechanism, where the total probability of some modes decreases, while for other modes they increase by the same amount.

The development of the source/sink methods implies in a probability transfer mechanism, with integrated modes probabilities $P_i(t)$ that depend on time, but conserving the total probability constant. There are many ways to build such a source/sink model by defining a transfer matrix
${\cal M}_{ij}(x,t)$, with $i,j=0,1,...$, such that 
\begin{equation}
 \begin{cases}
 \displaystyle
   \sum_{i,j} \frac{d  {\cal M}_{ij}}{dt}=0 \\
   \displaystyle
   \sum_{ij}\int dx {\cal M}_{ij}=0
 \end{cases}.\label{eq:transfermatrix}
\end{equation}
The matrix ${\cal M}$ can be general, but must account for the unmatched monomials in the PPE.

Integrating the PPE over all $x$ for mode $i$, one has
\begin{equation}
\int \frac{\partial f_i}{\partial t} dx = \int \frac{\partial}{\partial x}(k x^\alpha f_i) dx + D \int \frac{\partial^2}{\partial x^2}(f_i^{2-q}) dx. \label{eq:IntegralPPE}
\end{equation}
The integrated drift contribution vanishes identically,
\begin{equation}
\int_{-\infty}^{+\infty} \frac{\partial}{\partial x}\!\bigl[A(x)f_i(x,t)\bigr]\,dx
= \Bigl[A(x)f_i(x,t)\Bigr]_{-\infty}^{+\infty}
= 0,
\end{equation}
because the flux $A(x)f_i(x,t)$ decays faster than any power of $|x|$ at infinity for all admissible $q$ and $\alpha$ that make the stationary solution normalisable. Therefore, the unmatched monomials must be compensated, considering a time-dependent probability transfer that includes only the diffusion terms.

This is accomplished by adding to the probability density a time-independent function $F(t)$, such that $dF/dt=0$. $F(t)$ is associated with the transfer matrix by
\begin{equation}
 F(t)=\sum_{ij} M_{ij},
\end{equation}
hereby satisfying the necessary condition of time-independency due to Eq.~(\ref{eq:transfermatrix}).

To define a transfer matrix that tackles the unmatched monomials from the diffusion term, the mode probability
\begin{equation}
P_i(t) = N_i(t) \sigma_i(t) \mathcal{J}_q(\beta_i),
\qquad
\mathcal{J}_q(\beta_i)=2\int_0^{+\infty}\!\!
 \bigl[1+(q-1)w^{\beta_i}\bigr]^{-\frac{1}{q-1}}\,dw, \label{eq:DefPi}
\end{equation}
is introduced. Since $\sigma_i = \text{const}$ for $i \geq 1$, $\mathcal{J}_q(\beta_i) = \text{const}$, it results $\dot P_i = \dot N_i \sigma_i \mathcal{J}_q(\beta_i)$. Thus, the probability transfer is directly associated with the mode's amplitude $N_i$. The transfer matrix can be fully determined by
\begin{equation}
 {\cal M}_{ij}=
 \begin{cases}
  \dot P_i \quad \text{for $j=0$, $\forall i \ne 0$} \\
  0 \quad \text{for $i=0$, $\forall j$} \\
  \dot P_0=-\sum \dot P_i \quad \text{for $i=0$, $j=0$}
 \end{cases}
\end{equation}

From the diffusion term, it results
\begin{align}
D(q-2) \int \Bigl[& c_i\beta_i(\beta_i-1)|z_i|^{\beta_i-2}\frac{f_i}{N_i^{q-1}} \notag \\
& - c_i^2\beta_i^2|z_i|^{2\beta_i-2}\frac{[f_i]^q}{N_i^{2(q-1)}} \Bigr] dx.
\end{align}
The second derivative in the diffusion term yields
\begin{align}
\frac{\partial^2}{\partial x^2} (f_i^{2-q})
&= D(q-2)\Bigl[
c_i \beta_i (\beta_i-1) |z_i|^{\beta_i-2} \frac{f_i}{N_i^{q-1}}
- c_i^2 \beta_i^2 |z_i|^{2\beta_i-2} \frac{f_i^q}{N_i^{2(q-1)}}
\Bigr].
\end{align}
Integrating term by term, therefore, requires only the two spatial moments
\begin{align}
\mathcal{M}_1^{(i)} &= \int_{-\infty}^{+\infty} |z_i|^{\beta_i-2}\,f_i(x,t)\,dx, \\
\mathcal{M}_2^{(i)} &= \int_{-\infty}^{+\infty} |z_i|^{2\beta_i-2}\,f_i(x,t)^q\,dx.
\end{align}

These moments are evaluated exactly using the Euler Beta function, giving
\begin{align}
\mathcal{M}_1^{(i)}
&= \frac{2\,N_i(t)\,\sigma_i^{\beta_i-1}}{\beta_i}\;
   B\!\left(\frac{1}{\beta_i},\;\frac{1}{q-1}-\frac{1}{\beta_i}\right),
\label{eq:M1_beta}\\
\mathcal{M}_2^{(i)}
&= \frac{2\,N_i(t)^q\,\sigma_i^{2\beta_i-1}}{\beta_i}\;
   B\!\left(\frac{2}{\beta_i},\;\frac{q}{q-1}-\frac{2}{\beta_i}\right).
\label{eq:M2_beta}
\end{align}

These expressions are valid throughout the physical region where the stationary solution is normalisable [Eq.~\eqref{eq:conditions}].  
They automatically reduce to the familiar scale-invariant integrals when $\beta_i=2$ (q-Gaussian case):
$$
\mathcal{M}_1^{(i)}\to\sigma_i\,\mathcal{J}_q(2),\qquad 
\mathcal{M}_2^{(i)}\to N_i^{q-1}\sigma_i^{3}\,\mathcal{J}_q(4).
$$
Inserting Eqs.~\eqref{eq:M1_beta}--\eqref{eq:M2_beta} into the integrated diffusion term and using Eq.~(\ref{eq:DefPi}) yields
\begin{equation}
\dot{P}_i(t) = A_i - B_i\,P_i^{q-2}(t), \label{eq:Pt}
\end{equation}
with time-independent coefficients
\begin{align}
A_i &= D(q-2)\,(\beta_i-1)\,\beta_i\,\sigma_i^{-\beta_i}\;
       B\!\left(\frac{1}{\beta_i},\;\frac{1}{q-1}-\frac{1}{\beta_i}\right),\\
B_i &= D\,|q-2|\,\beta_i^{2}\,\sigma_i^{-2\beta_i}\;
       \frac{B\!\left(\frac{2}{\beta_i},\;\frac{q}{q-1}-\frac{2}{\beta_i}\right)}
            {B\!\left(\frac{1}{\beta_i},\;\frac{1}{q-1}-\frac{1}{\beta_i}\right)}.
\end{align}
The solution of Eq.~(\ref{eq:Pt}) can be written in the form
\begin{equation}
    t-t_0=\int_{P_i(t_0)}^{P_i(t)} \frac{d P'_i}{A_i-B_i (P'_i)^{q-2}},
\end{equation}
which can be expressed in terms of the Lerch transcendent function, $\Phi$,
\begin{equation}
t - t_0 = \frac{P_i(t)^{3-q}}{(3-q) A_i} \, \Phi\left(\frac{B_i}{A_i} P_i(t)^{q-2}, 1, \frac{3-q}{q-2}\right) - \frac{P_i(t_0)^{3-q}}{(3-q) A_i} \, \Phi\left(\frac{B_i}{A_i} P_i(t_0)^{q-2}, 1, \frac{3-q}{q-2}\right)
\end{equation}

However, for the relevant cases where the PPE can be used, $A_i$ will be negative and $|A_i| \gg B_i$, hence, to a good approximation, one can write
\begin{equation}
    P_i(t)=\max[P_i(0)-v_i t,0], \qquad v_i=-A_i .
\end{equation}
Otherwise, in applications where the $A_i \ll B_i P_i^{q-2}(t)$, the approximation
\begin{equation}
\dot{P}_i(t) \approx -\Gamma_i^{(0)}\,P_i^{q-2}(t),\qquad i\ge 1,
\end{equation}
with the exact decay rate
\begin{equation}
\Gamma_i^{(0)} 
= D\,|q-2|\;\beta_i^{2}\sigma_i^{-2\beta_i}\,.
\end{equation}

\subsection{Attractor mode}

Total probability conservation forces:
\begin{equation}
\dot P_0(t) = \sum_{i=1}^M \Gamma_i^{(0)} P_i^{q-2}(t).
\end{equation}

For the attractor mode ($i=0$), $\beta_0 = \alpha + 1$ is fixed, and the stationary relation holds locally:
\begin{equation}
N_0^{q-1} = \frac{D(q-2)(\alpha+1)}{k} = \kappa.
\end{equation}
Thus,
\begin{equation}
N_0(t) = \kappa^{1/(q-1)} = \text{constant}.
\end{equation}

The probability content is:
\begin{equation}
P_0(t) = N_0 \sigma_0(t) \mathcal{J}_q(\alpha + 1).
\end{equation}

Differentiating with respect to time:
\begin{equation}
\dot P_0(t) = N_0 \mathcal{J}_q(\alpha + 1) \dot{\sigma}_0(t).
\end{equation}

Substituting the expression for $\dot P_0(t)$:
\begin{equation}
\dot{\sigma}_0(t) 
  = \frac{1}{N_0 \, \mathcal{J}_q(\alpha+1)} 
    \sum_{i=1}^{M} \Gamma_i^{(0)} P_i^{q-2}(t),
\qquad
\sigma_0(t) 
  = \sigma_0(0) 
    + \frac{1}{N_0 \, \mathcal{J}_q(\alpha+1)} 
      \sum_{i=1}^{M} \bigl[ P_i(0) - P_i(t) \bigr].
\label{eq:sigma0_dot}
\end{equation}
This shows that the attractor width evolves in time to accommodate the incoming probability flux, while all transient mode widths remain constant.
Remarkably, when $\alpha=1$ (linear drift), the attractor width evolves as $\sigma_0(t) \propto  t^{q-1} $, which is proportional to the transferred probability, exactly recovering the well-known spreading law of single-mode q-Gaussian solutions — now valid for arbitrary multimodal initial conditions. 

The table below summarises the complete set of coupled equations for the multi-mode approach:
\begin{equation}
\boxed{
\begin{aligned}
&\dot{x}_{0,i}(t) = -k \, [x_{0,i}(t)]^{\alpha}, 
&& i=0,\dots,M, \\[3pt]
&\beta_0 = \alpha + 1 \quad \text{(fixed)}, \quad
 \beta_i = \text{const} \; (i\ge 1),\;\;
 \beta_i \neq \beta_j\;(i \neq j), \\[3pt]
&A_i = D(q-2)\,(\beta_i-1)\,\beta_i\,\sigma_i^{-\beta_i}
       B\!\left(\frac{1}{\beta_i},\;\frac{1}{q-1}-\frac{1}{\beta_i}\right),\\[3pt]
&B_i = D\,|q-2|\,\beta_i^{2}\,\sigma_i^{-2\beta_i}\;
       \frac{B\!\left(\frac{2}{\beta_i},\;\frac{q}{q-1}-\frac{2}{\beta_i}\right)}
            {B\!\left(\frac{1}{\beta_i},\;\frac{1}{q-1}-\frac{1}{\beta_i}\right)} \\[3pt]
&P_i(t) = \max{P_i(0) +A_i t,0}, 
&& A_i \gg B_i \\[3pt]
&P_i(t) = P_i(0) \Bigl[1 + (q-1) \frac{t}{\tau_i}\Bigr]^{-\frac{3-2q}{q-1}}=P_i(0) e_q(t/\tau_i)^{2q-3}, 
&& A_i \ll B_i \\[3pt]
&{P}_0(t) = P_0(0)+\sum_i P_i(0)-P_i(t) \bigr]^{-1} && i=1,\dots,M, \\[4pt]
&\dot{\sigma}_i(t) = 0, 
&& i=1,\dots,M, \\[3pt]
&\sigma_0(t)
  = \sigma_0(0)
    + \frac{\sum_{i=1}^{M} \bigl[ P_i(0) - P_i(t) \bigr]}{N_0 \, \mathcal{J}_q(\alpha+1)}, \\[3pt]
&N_0 = \left[\frac{D(q-2)(\alpha+1)}{k}\right]^{\!1/(q-1)}, 
\quad N_i(t) = \frac{P_i(t)}{\sigma_i \, \mathcal{J}_q(\beta_i)}, 
&& i=0,\dots,M.
\end{aligned}}
\label{eq:coupled_system_complete}
\end{equation}

\subsection{Special Cases for Specific Drift Exponents}

The multi-mode framework allows both the attractor centre $x_{0,0}(t)$ and width $\sigma_0(t)$ to vary with time for arbitrary $\alpha$, provided there are transient modes contributing probability flux (i.e., $M \geq 1$). The evolution of $x_{0,0}(t)$ depends explicitly on $\alpha$ through Eq.~(\ref{eq:x0i_dot}), while $\sigma_0(t)$ evolves via Eq.~(\ref{eq:sigma0_dot}), independent of $\alpha$ but scaled by $\mathcal{J}_q(\beta_0 = \alpha + 1)$. Below, we highlight special cases for $\alpha=1$ (linear drift), $\alpha=0$ (constant drift), and $\alpha=-0.5$ (fractional drift), assuming the normalizability conditions in Eq.~(\ref{eq:conditions}) hold (e.g., $1 < q < 2$, $\alpha + 1 > 0$, $k D > 0$).

In all cases, transient widths are constant ($\dot{\sigma}_i(t) = 0$ for $i \geq 1$), and transient probabilities decay as $P_i(t) = P_i(0) \left[1 + (q-1) \frac{t}{\tau_i}\right]^{\frac{3-2q}{q-1}}$, with $\tau_i = \left[ \Gamma_i^{(0)} P_i(0)^{q-2} \right]^{-1}$ and $\Gamma_i^{(0)}$ given by Eq.~(\ref{eq:Gamma_beta}). The stationary relation fixes the attractor amplitude $N_0$.

\subsubsection{Linear Drift ($\alpha = 1$, $\beta_0 = 2$): q-Gaussian Attractor}

For linear drift $A(x) = -k x$ ($k > 0$ for restoring), the attractor is a q-Gaussian profile. The centre evolves exponentially:
\begin{equation}
\dot{x}_{0,0}(t) = -k x_{0,0}(t) \quad \Rightarrow \quad x_{0,0}(t) = x_{0,0}(0) \, e^{-k t},
\end{equation}
approaching the equilibrium at $x=0$. The width grows proportionally to the transferred probability:
\begin{equation}
\sigma_0(t) = \sigma_0(0) + \frac{1}{N_0 \, \mathcal{J}_q(2)} \sum_{i=1}^{M} \left[ P_i(0) - P_i(t) \right].
\end{equation}
If transients share similar $\beta_i$ and thus $\tau_i \approx \tau$, $\sigma_0(t) \propto 1 - \left[1 + (q-1) \frac{t}{\tau}\right]^{\frac{3-2q}{q-1}}$. For $1 < q < 2$, this yields initial power-law growth in $t$, saturating at long times. This recovers known single-mode q-Gaussian spreading laws for arbitrary initial conditions decomposable into modes with $\beta_i \neq 2$.

\subsubsection{Constant Drift ($\alpha = 0$, $\beta_0 = 1$): q-Exponential Attractor}

For constant drift $A(x) = -k$ (uniform force), the attractor is a pure q-exponential. The centre moves linearly:
\begin{equation}
\dot{x}_{0,0}(t) = -k \quad \Rightarrow \quad x_{0,0}(t) = x_{0,0}(0) - k t.
\end{equation}
The width evolves as:
\begin{equation}
\sigma_0(t) = \sigma_0(0) + \frac{1}{N_0 \, \mathcal{J}_q(1)} \sum_{i=1}^{M} \left[ P_i(0) - P_i(t) \right],
\end{equation}
with the same power-law form driven by transients with $\beta_i \neq 1$. This extends known single-mode travelling-wave solutions (constant $\sigma_0$, varying $x_{0,0}$) to include time-varying widths, useful for modelling biased diffusion in fractal media or urban redistribution under constant external forces.

\subsubsection{Fractional Drift ($\alpha = -0.5$, $\beta_0 = 0.5$): Sublinear Attractor}

For sublinear drift $A(x) = -k x^{-0.5}$ (admissible for $\beta_0 = 0.5 > 0$), the centre evolves via:
\begin{equation}
\dot{x}_{0,0}(t) = -k [x_{0,0}(t)]^{-0.5} \quad \Rightarrow \quad \int_{x_{0,0}(0)}^{x_{0,0}(t)} x^{0.5} \, dx = -k t,
\end{equation}
yielding $x_{0,0}(t) = \left[ x_{0,0}(0)^{1.5} - \frac{3}{2} k t \right]^{2/3}$ (for $x_{0,0} > 0$, assuming restoring behavior). The width follows:
\begin{equation}
\sigma_0(t) = \sigma_0(0) + \frac{1}{N_0 \, \mathcal{J}_q(0.5)} \sum_{i=1}^{M} \left[ P_i(0) - P_i(t) \right],
\end{equation}
again with power-law transients. This case is relevant for systems with weakening drift at large $|x|$, such as certain fractal or nonextensive models, where no exact single-mode time-dependent solutions exist, but multi-mode solutions provide an analytic approximation.

\section{Application: Relaxation of linearly weighted q-Gaussian Distributions in Fractal Socioeconomic Networks}
\label{sec:application_urban}

A complete example of the multi-mode framework application is inspired by the modelling of socioeconomic variables on fractal social networks, where q-Gaussian profiles arise naturally from hierarchical interactions~\citep{Deppman2025_FractalConsensus}. In such systems, the observable output (wealth, productivity, influence, innovation, etc.) is often proportional to an underlying continuous variable $x$ (skill, connectivity, cumulated wealth, etc.). For instance, the potential to generate wealth is often an increasing function of the cumulated wealth. For a hypothetical example, the generated wealth follows a linear relation with the individual's cumulated wealth. Thus, at a given time, the wealth per individual is not a $q$-Gaussian distribution, but an effective initial distribution that has the form of a linear-weighted q-Gaussian,
\begin{equation}
\tilde{f}(x,0) \propto \max\!\Bigl\{a + b x,\,0\Bigr\}\,
e_q^+\!\left(\frac{x^2}{2\sigma^2}\right).
\label{eq:linear_weighted}
\end{equation}

Although Eq.~\eqref{eq:linear_weighted} is not a finite sum of the pure power-law modes of Eq.~\eqref{eq:ansatz_independent}, it can be expanded (numerically or analytically via a multipole-like decomposition) in the complete q-exponential basis with independent centres $x_{0,i}(t)$ and arbitrary exponents $\beta_i$. For linear drift $A(x)=-k x$ ($\alpha=1$) the attractor is the usual q-Gaussian ($\beta_0=\alpha+1=2$), centred at the origin with fixed amplitude
\begin{equation}
N_0 = \left[\frac{D(2-q)\cdot 2}{|k|}\right]^{\frac{1}{q-1}}.
\end{equation}
This function can be expanded in terms of a series of stretched $q$-exponentials, as illustrated in Fig.~\ref{fig:Multimode-example}, where the black line represents the initial distribution (a linear-weighted $q$-Gaussian) and the dark-blue curve is the fitted series of eight stretched $q$-exponentials with parameters given in Table~\ref{tab:strong_asym}. With such an expansion, the multi-mode procedure can be easily applied.

Under the exact dynamics derived in Section~\ref{sec:multi} one obtains:
\begin{itemize}
\item \emph{Centre motion} (consensus formation):
  \begin{equation}
  \dot{x}_{0,i}(t) = -k x_{0,i}(t) \;\Rightarrow\; x_{0,i}(t) = x_{0,i}(0)\,e^{-k t}.
  \end{equation}
  For confining drift $k>0$ the centres of every transient pair exponentially approach the consensus point $x=0$ at a common rate.

\item \emph{Constant transient widths}: $\dot{\sigma}_i(t)=0$ for all $i\ge 1$.

\item \emph{Power-law relaxation of transient probability content}:
  \begin{equation}
  P_{\pm i}(t) = P_i(0)\left[1-(2-q)\Gamma_i^{(0)}P_i(0)^{q-2}t\right]^{\frac{1}{2-q}}
  \end{equation}
  with mode-dependent rates $\Gamma_i^{(0)} \propto \beta_i^2/\sigma_i^{2\beta_i}$ (multi-scale relaxation).

\item \emph{Irreversible absorption by the attractor}: the total flux
  \begin{equation}
  \dot{P}_0(t) = \sum_{i\ge1}\Gamma_i^{(0)}P_i^{q-2}(t)
  \end{equation}
  forces monotonic growth of the attractor width
  \begin{equation}
  \sigma_0(t) = \sigma_0(0)\left[1+\frac{q-1}{\mathcal{J}_q(2)}\int_0^t\!\sum_{i\ge1}\Gamma_i^{(0)}P_i(t')^{3-q}\,dt'\right]^{\frac{1}{q-1}},
  \end{equation}
  while $N_0$ remains constant.
\end{itemize}

Fig.~\ref{fig:Mode-Evolution} displays the evolution of the individual modes (grey lines) and of the total distribution (black line) at the initial instances. It is observed that an initially skewed, linear-weighted q-Gaussian relaxes toward the symmetric stationary q-Gaussian through: (i) drift of the modes toward the centre of attraction (origin); (ii) power-law (non-Markovian) decay of modal amplitudes; and (iii) irreversible broadening of the attractor to accommodate the incoming probability (increasing variance and persistent inequality).

This mechanism provides the first exact analytic description of the simultaneous emergence of consensus and inequality in fractal hierarchical societies: the linear weighting encodes initial socioeconomic bias, the linear drift drives exponential convergence to a common centre, and the nonextensive nonlinear diffusion produces the heavy-tailed, multi-scale redistribution dynamics universally observed in wealth, innovation, and urban indicators \citep{Bettencourt2013,Deppman2024_Urban,Deppman2025_FractalConsensus}. The same mathematics applies to transient skewness in heavy-quark jet distributions before full thermalisation in quark--gluon plasma and to post-catastrophe recovery of urban population profiles toward the universal q-Gaussian attractor.

\begin{figure}[htbp]
    \centering
    \includegraphics[width=0.77\linewidth]{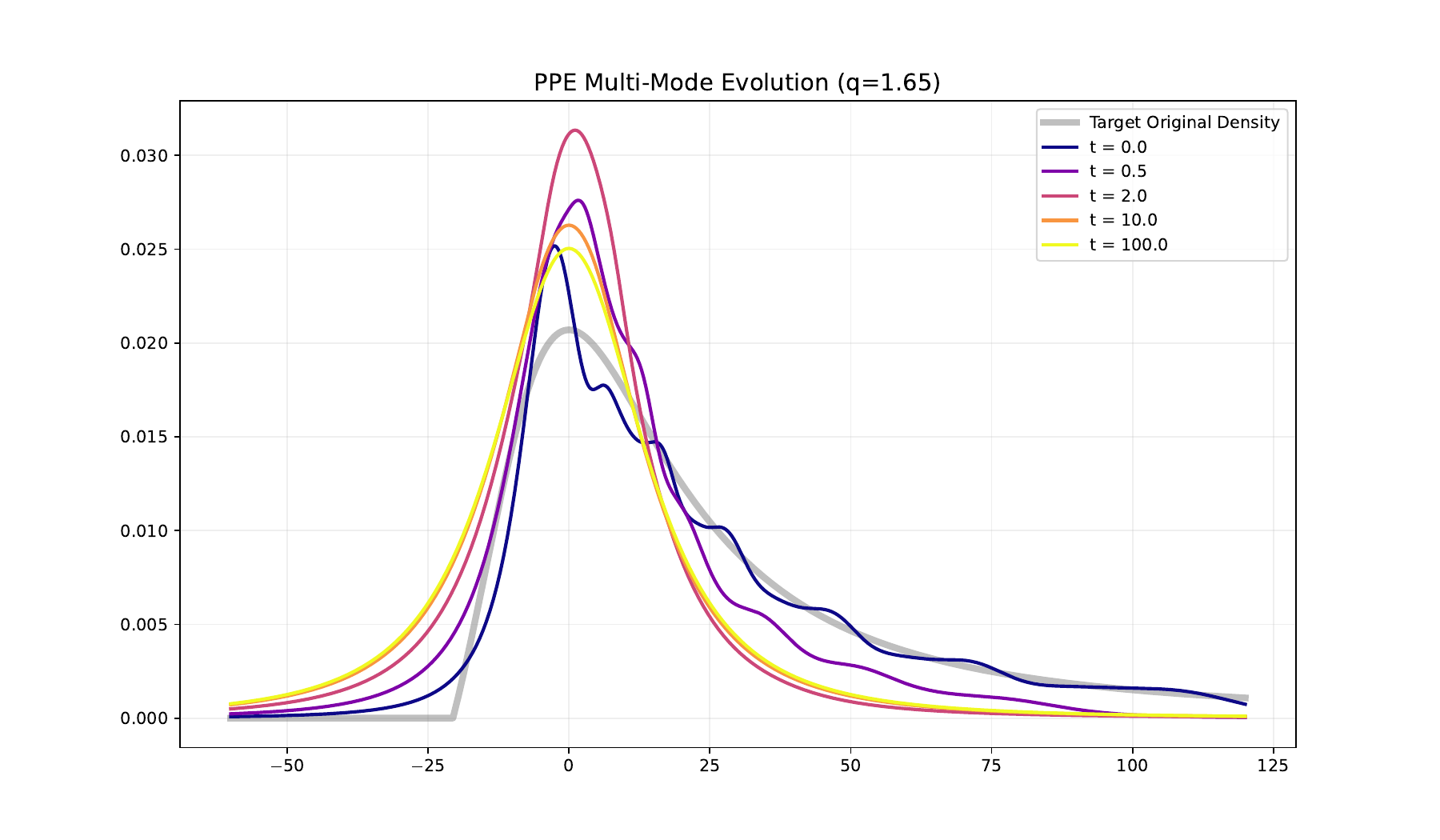}
    \caption{Example of application of the multi-modal solution. The hypothetical case of a socioeconomic output (wealth, skill, social influence, opinion, etc.) in an initially distorted probability density (blue line at $t=0$), which evolves in time reaching, after some time (in arbitrary units), the stationary regime in red ($t=32$). The example uses six modes: the attractor (mode 0) and five transient modes. The fit of the initial distribution (target) can be improved by including additional modes. See the text for details on the parameters used.}
    \label{fig:Multimode-example}
\end{figure}

\begin{table}[ht]
\centering
\caption{Initial multi-mode fit parameters ($t=0$) for $q=1.65$ and $\beta_{attr}=2.0$. The parameters represent the attractor (mode number 0) and transient $q$-exponential modes used to reconstruct the target density.}
\label{tab:initial_params}
\begin{tabular}{ccccc}
\hline
Mode ($i$) &  Amplitude ($A_i$) & Center ($x_{0,i}$) & Width ($\sigma_i$) & Shape ($\beta_i$) \\
\hline
0 &  $3.2851 \times 10^{-1}$ & $-2.6176$ & $8.1508$ & $2.0000$ \\
1  & $1.1296 \times 10^{-1}$ & $24.7535$ & $7.5776$ & $3.1309$ \\
2 &  $1.0094 \times 10^{-1}$ & $41.4126$ & $11.0894$ & $3.5393$ \\
3 &  $6.8190 \times 10^{-2}$ & $64.9444$ & $14.0474$ & $3.4107$ \\
4 &  $1.4270 \times 10^{-1}$ & $11.5591$ & $7.4194$ & $4.9360$ \\
5 &  $5.8323 \times 10^{-2}$ & $98.8087$ & $21.4152$ & $3.3769$ \\
\hline
\end{tabular}
\end{table}

\begin{figure}[htbp]
    \centering
    \begin{subfigure}[b]{0.47\linewidth} 
      \includegraphics[width=\linewidth]{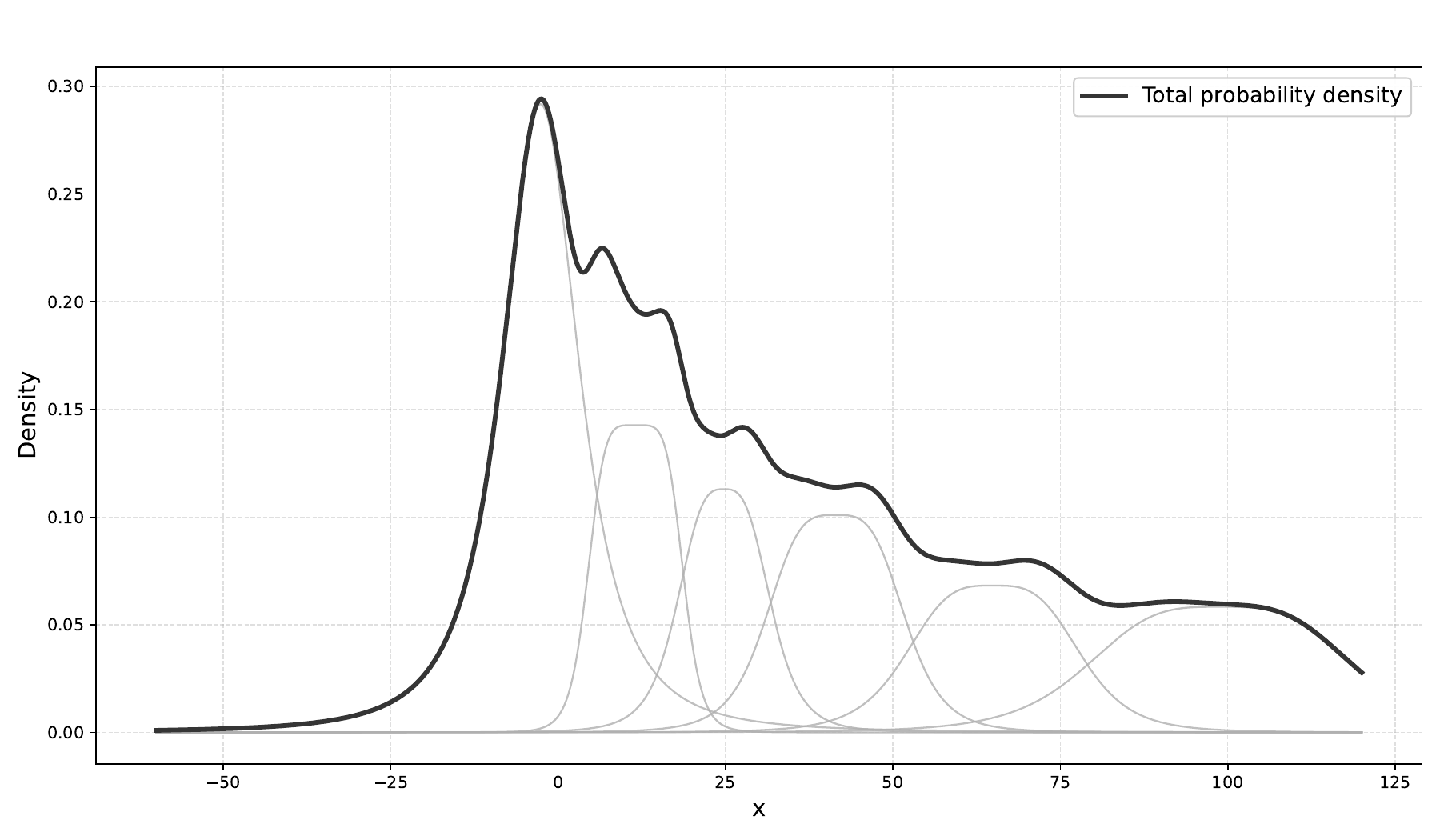}
    \end{subfigure}
    \begin{subfigure}[b]{0.47\linewidth} 

    \includegraphics[width=\linewidth]{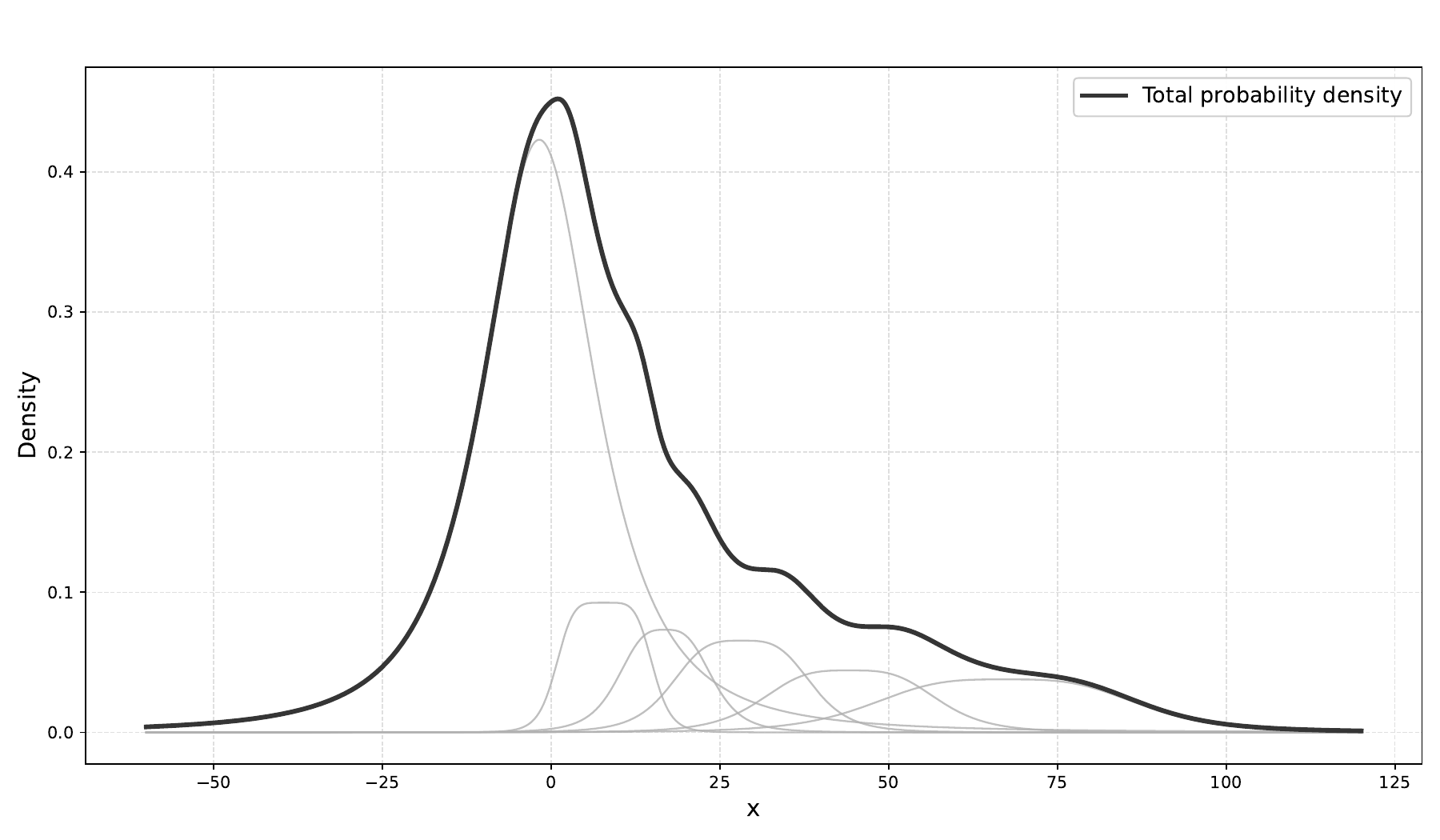}
    \end{subfigure}
    \begin{subfigure}[b]{0.47\linewidth} 
      \includegraphics[width=\linewidth]{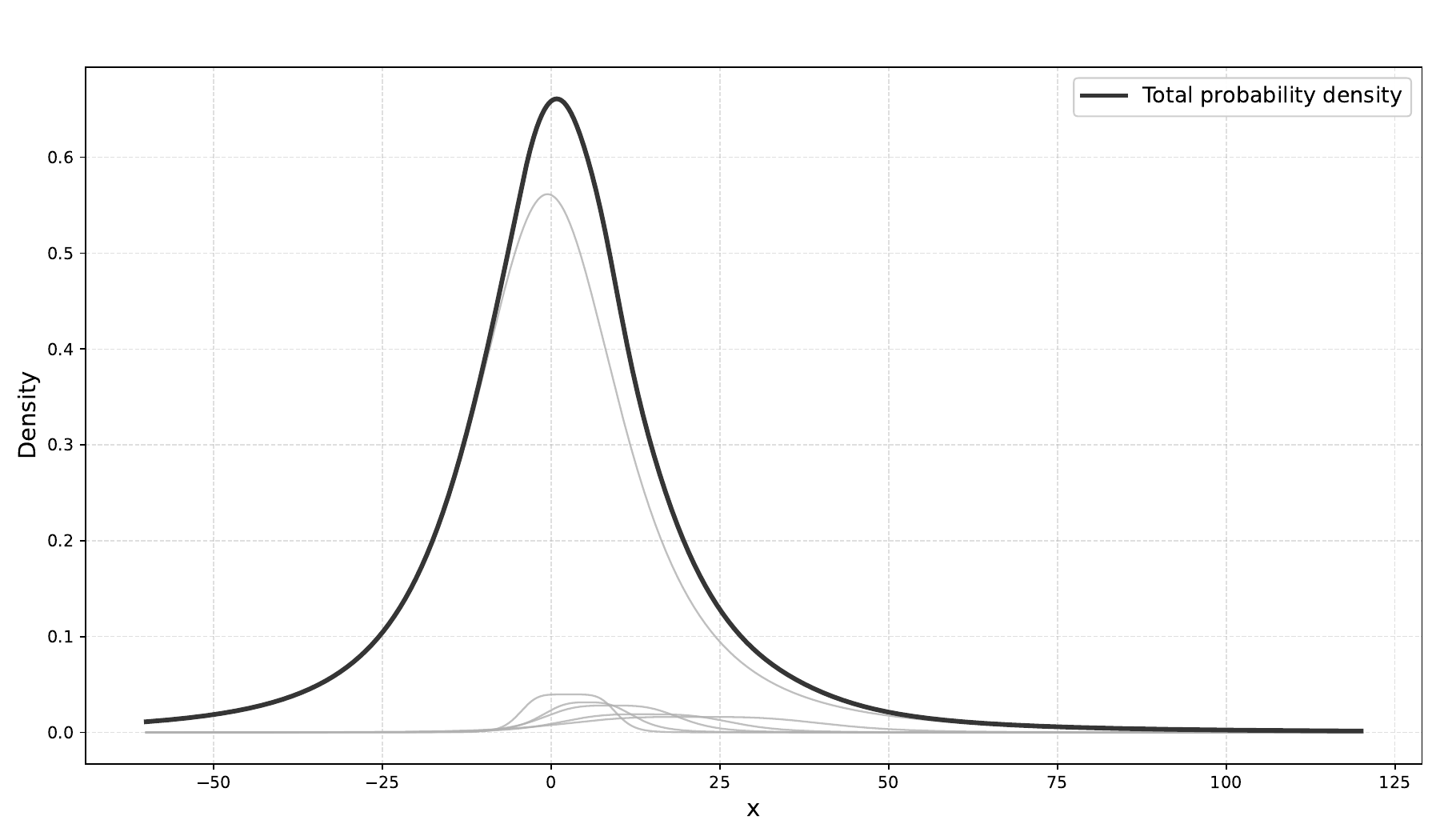}
    \end{subfigure}
    \begin{subfigure}[b]{0.47\linewidth} 
      \includegraphics[width=\linewidth]{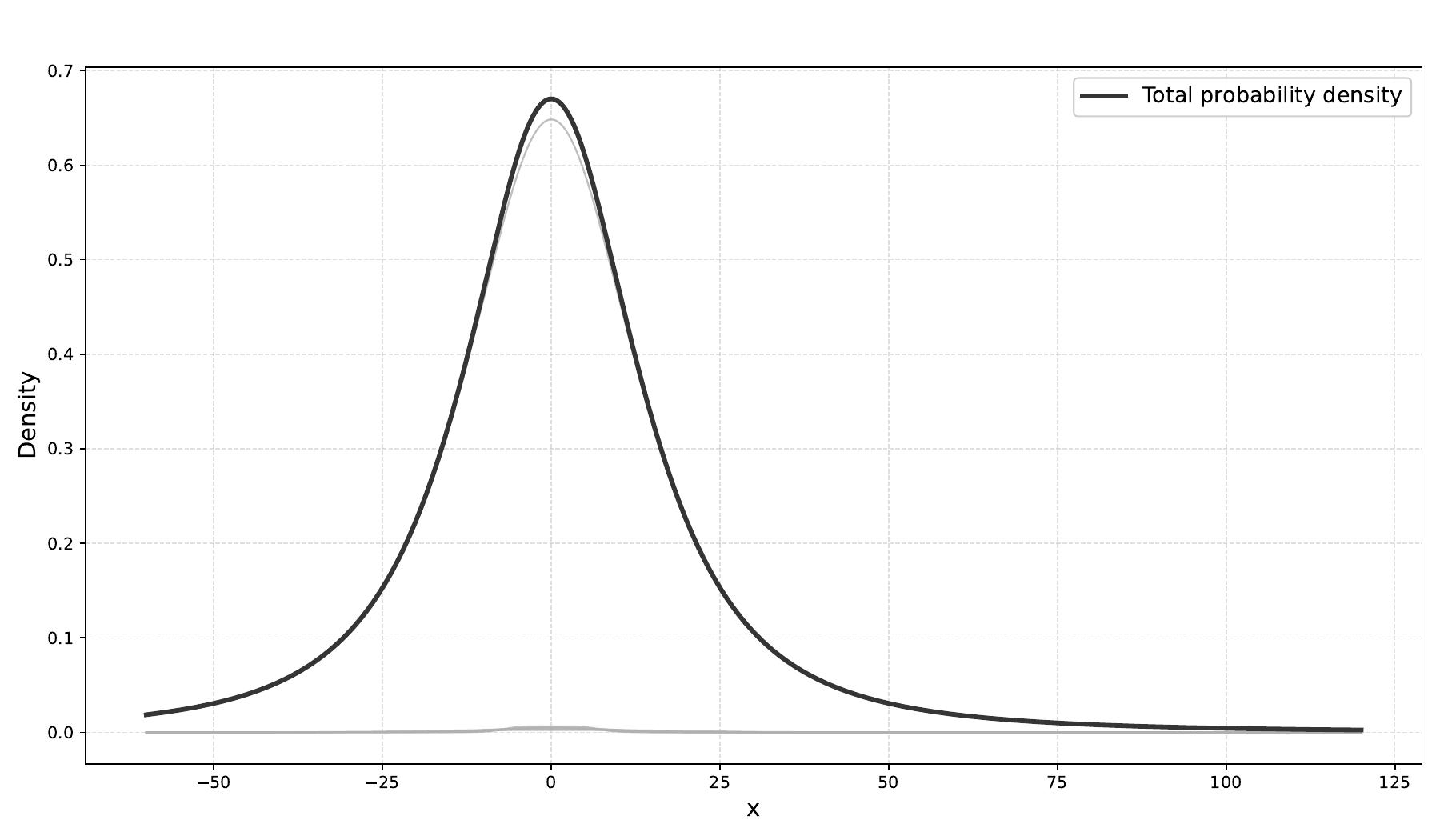}
    \end{subfigure}
    \caption{Time evolution of the modes and of the total probability density. The dynamical evolution time (arbitrary units) is: $t=0.0$ (top-left); $t=0.5$ (top-right); $t=2.0$ (bottom-left); and $t=10.0$ (bottom-right). The darker line represents the sum of the components, which are shown in the grey lines.}
    \label{fig:Mode-Evolution}
\end{figure}

\section{Conclusions}
\label{sec:conclusions}

We have constructed the first complete, exact, multi-mode family of solutions to the Plastino--Plastino equation for arbitrary power-law drift $A(x)=-kx^\alpha$. The crucial innovation, allowing each mode to possess its own time-dependent centre $x_{0,i}(t)$ eliminates cross-mode coupling in the nonlinear drift term and leads to an unprecedented degree of analytical tractability: the infinite hierarchy of moments closes exactly, widths of all transient modes remain strictly constant, centre motion decouples mode-by-mode, and probability content evolves according to fully separable ordinary differential equations exhibiting exact q-exponential (power-law) relaxation.

A single attractor mode with $\beta_0=\alpha+1$ plays a privileged role: its amplitude is fixed by the transport coefficients, while its width $\sigma_0(t)$ grows in time to absorb the entire probability flux released by the decaying transient modes, thereby driving the system irreversibly toward the known stationary q-exponential profile from essentially arbitrary initial conditions decomposable in the q-exponential basis. All previous exact solutions (linear drift, harmonic drift, single-mode travelling waves) emerge as special or limiting cases of the present framework.

These results significantly extend the analytical reach of nonextensive statistical mechanics and nonlinear diffusion theory. They provide, for the first time, a transparent and exact description of multi-scale relaxation dynamics in systems governed by Tsallis entropy: how transient structures decay, how probability is transferred across scales, and how the universal q-exponential attractor is approached in finite time. The separability of centre motion, the constancy of transient widths, and the exact power-law form of $P_i(t)$ constitute sharp, parameter-free predictions that can be directly tested against high-precision numerical simulations and, more importantly, against experimental and observational data in a wide variety of complex systems. Crucially, the choice of fixed-width transient modes eliminates the most dangerous nonlinear couplings and is the key technical advance that allows exact closure of the hierarchy.

The applications of the formalism presented here are immediate and far-reaching. In fractal media (granular materials, biological tissues, turbulent flows), they capture Lévy-flight-like multimodal transients on self-similar substrates. In urban science, they model post-disruption population redistribution across cities or neighbourhoods after catastrophic events, with each mode representing a displaced subpopulation drifting under effective gravitational-like forces. In quark--gluon plasma physics, the solutions describe the simultaneous evolution of multiple heavy-quark jets or fireballs with different initial rapidities, offering an exact benchmark for extracting genuine nonextensive signatures from LHC data.

Extensions to higher dimensions, inhomogeneous diffusion coefficients, and coupled multi-species systems are now analytically accessible. Future work will focus on exploring the possibility of a self-organised centre of attraction, a feature that can be useful to understand complex dynamics, opening a way to further investigate how the combination of drift and diffusion transport coefficients can alter the dynamics leading to different stationary states. Another important line of future research lies in exploring the thermodynamic and information-geometric interpretation of probability transfer between modes. In addition, confronting the predicted multi-mode relaxation spectra with high-resolution data from heavy-ion collisions, living tissues, and urban recovery processes will clarify all the potentialities of the framework developed in this work.

By unifying fractal derivatives, q-deformed calculus, and nonlinear Fokker--Planck dynamics into a single exactly solvable framework, the present solutions reinforce the status of the Plastino--Plastino equation as the natural evolution equation for complex systems with power-law stationary states, long-range correlations, and scale-free environments, from quark--gluon plasma to human cities.

 \bibliographystyle{elsarticle-num}

\begin{thebibliography}{10}
\expandafter\ifx\csname url\endcsname\relax
  \def\url#1{\texttt{#1}}\fi
\expandafter\ifx\csname urlprefix\endcsname\relax\def\urlprefix{URL }\fi
\expandafter\ifx\csname href\endcsname\relax
  \def\href#1#2{#2} \def\path#1{#1}\fi

\bibitem{Plastino1995}
A.~R. Plastino, A.~Plastino, Non-extensive statistical mechanics and generalized {F}okker-{P}lanck equation, Physica A: Statistical Mechanics and its Applications 222~(1-4) (1995) 347--354.
\newblock \href {https://doi.org/10.1016/0378-4371(95)00243-4} {\path{doi:10.1016/0378-4371(95)00243-4}}.

\bibitem{Nobre2019}
G.~A. Casas, F.~D. Nobre, Nonlinear fokker-planck equations in super-diffusive and sub-diffusive regimes, Journal of Mathematical Physics 60 (2019).

\bibitem{Tsallis1988}
C.~Tsallis, Possible generalization of {B}oltzmann-{G}ibbs statistics, Journal of Statistical Physics 52~(1-2) (1988) 479--487.
\newblock \href {https://doi.org/10.1007/BF01016429} {\path{doi:10.1007/BF01016429}}.

\bibitem{GellMann2004}
M.~Gell-Mann, C.~Tsallis (Eds.), Nonextensive Entropy: Interdisciplinary Applications, Oxford University Press, New York, 2004.
\newblock \href {https://doi.org/10.1093/acprof:oso/9780195159769.001.0001} {\path{doi:10.1093/acprof:oso/9780195159769.001.0001}}.

\bibitem{Tsallis2023}
C.~Tsallis, Non-additive entropies and statistical mechanics at the edge of chaos: a bridge between natural and social sciences, Philosophical Transactions of the Royal Society A: Mathematical, Physical and Engineering Sciences 381 (2023).

\bibitem{Deppman2023}
A.~Deppman, E.~Megías, R.~Pasechnik, Fractal derivatives, fractional derivatives and q-deformed calculus, Entropy 25 (2023) 1008.

\bibitem{Golmankhaneh2019}
A.~K. Golmankhaneh, C.~Tun\c{c}, Stochastic differential equations on fractal sets, Stochastics 92 (2019) 1244–1260.

\bibitem{Megas2024}
E.~Megías, A.~Khalili~Golmankhaneh, A.~Deppman, Dynamics in fractal spaces, Physics Letters B 848 (2024) 138370.

\bibitem{Deppman2023b}
A.~Deppman, A.~Khalili~Golmankhaneh, E.~Megías, R.~Pasechnik, {From the Boltzmann equation with non-local correlations to a standard non-linear Fokker-Planck equation}, Physics Letters B 839 (2023) 137752.

\bibitem{Megias2023_Comparative}
E.~Megias, A.~Deppman, R.~Pasechnik, C.~Tsallis, Comparative study of the heavy-quark dynamics with the {F}okker-{P}lanck equation and the {P}lastino-{P}lastino equation, Physics Letters B 845 (2023) 138136.

\bibitem{Tsallis1996}
C.~Tsallis, D.~J. Bukman, Anomalous diffusion in the presence of external forces: Exact time-dependent solutions and their thermostatistical basis, Physical Review E 54~(3) (1996) R2197--R2200.

\bibitem{Combe2015}
G.~Combe, V.~Richefeu, M.~Stasiak, A.~P. Atman, Experimental validation of a nonextensive scaling law in confined granular media, Physical Review Letters 115 (2015).

\bibitem{Deppman2024_Urban}
A.~Deppman, R.~L. Fagundes, E.~Megias, R.~Pasechnik, F.~L. Ribeiro, C.~Tsallis, Dynamics of cities, Chaos, Solitons \& Fractals 191 (2024) 115877.

\bibitem{Borland2002}
L.~Borland, Option pricing formulas based on a non-{G}aussian stock price model, Physical Review Letters 89~(9) (2002) 098701.

\bibitem{PARVATE2009}
A.~Parvate, A.~D. Gangal, Calculus on fractal subsets of real line — {I}: Formulation, Fractals 17 (2009) 53–81.

\bibitem{PARVATE2011}
A.~Parvate, A.~D. Gangal, Calculus on fractal subsets of real line — ii: Conjugacy with ordinary calculus, Fractals 19 (2011) 271–290.

\bibitem{Batty1994}
M.~Batty, P.~Longley, Fractal Cities: A Geometry of Form and Function, Academic Press, London, 1994.

\bibitem{Deppman2025_Brain}
A.~Deppman, Urban scaling is hardwired in the human brain, British Journal of Multidisciplinary and Advanced Studies 6 (2025) 1–10.

\bibitem{Xie2025}
J.~Xie, R.~Ai, J.~Gao, F.~Liu, Y.~Tang, Motion and fixation: The urban evolution through the reconfiguration of scale structure, Cities 167 (2025) 106320.

\bibitem{Bettencourt2013}
L.~M.~A. Bettencourt, The origins of scaling in cities, Science 340~(6139) (2013) 1438--1441.

\end{thebibliography}

\end{document}